# Cavity Induced Tunable Extraordinary Transmission – A Unique Way of Funneling Light through Subwavelength Apertures

Sushrut Modak[1,3,‡], Alireza Safaei[2,3,‡], Debashis Chanda [1,2,3*]

[1]CREOL, The College of Optics and Photonics, University of Central Florida, Orlando, Florida 32826, USA.

[2]Department of Physics, University of Central Florida, Orlando, Florida 32826, USA.

[3]NanoScience Technology Center, University of Central Florida, Orlando, Florida 32826, USA.

[‡]These authors contributed equally.

**Extraordinary transmission through subwavelength metallic apertures has been extensively studied and demonstrated. At resonance, the coupling between surface plasmons on both surfaces of the metallic film tunnels the photon from the one side to the other through the subwavelength aperture with small transmission efficiency based on the metal's dielectric parameters and aperture geometrical dimensions. Here, we report a completely different form of extraordinary transmission, where the two subwavelength complementary apertures when coupled with an optical cavity induce about 100% extraordinary transmission far away from the natural plasmon resonance of the constituent metallic apertures. Such unique cavity phase driven plasmon resonance enables tuning of the extraordinary transmission band across wide spectral range unlike previously reported geometry driven extraordinary transmission.**

Diffraction does not allow light to pass through an aperture on an opaque film if the aperture dimension is sub-wavelength. This can be shown through diffraction analysis that constructive interference in the forward direction cannot take place for an aperture who's dimension is less than the wavelength of the incident light [1]. However, the situation changes if the incident light excites surface plasmon on the patterned film. The excitation of surface plasmons on the incident face of the film causes the free-space wavelength to shrink to that of the supported surface plasmon wavelength. The shortening of wavelength, coupling between plasmons on both sides of the film and re-radiation of oscillating electrons at resonance enables tunneling of the photon through the subwavelength aperture – called extraordinary light transmission [2-16]. Such extraordinary transmission has been widely studied in the context of single apertures [9,12,14,16] as well as array [3-5,7,8,10,11,13,15] in various shapes and geometry. The phenomenon solely depends on dielectric properties of the metal, aperture dimensions and periodicity which affects the transmission through near-field and far-field couplings [2,4-6,8,11,13-16].

Here, we report a completely different extraordinary transmission mechanism through subwavelength apertures. We demonstrate a cavity induced extraordinary transmission (CIET) where cavity phase determines a narrow transmission band ($\Delta\lambda/\lambda_0 < 0.075$) and transmitted energy through a subwavelength aperture ($\lambda \gg$ diameter(D)) unlike previously reported geometry driven extraordinary transmission. The same cavity-coupled hole-disk behaves as a multi-resonance system in the other two wavelength scale regimes ($\lambda \sim D$ and $\lambda \ll D$) as reported in our earlier work [17,18]. However, in the subwavelength regime ($\lambda \gg D$), the complimentary aperture pair, a hole and disk, when driven in-phase by the cavity, funnels ~100% of the incident photons through the subwavelength aperture. For a given metal and aperture dimension, the transmission band of which is widely tunable across a wide spectral range with the cavity phase unlike any other previously reported extraordinary transmission phenomenon. Further, from effective medium point of view, this cavity-coupled hole-disk array shows a near-zero effective refractive index at resonance due to the linear crossing of resonant electric and magnetic states as demonstrated in our work which will be reported in the future[19].

Figure 1a-inset schematically shows the complementary apertures - a coupled hole-disk array. The 30 nm thick subwavelength gold hole-array of diameter 0.76 µm and period of 1.14 µm has no extraordinary transmission over the 3-10 µm mid-IR band as can be observed from the FDTD prediction in Fig. 1a. The dimensions are far away from the resonant extraordinary transmission band of the hole-array. However, the addition of the complementary disk-array 74 nm below the perforated film increases the transmission to ~50% at $\lambda$ = 4.53 µm as can be seen in Fig. 1a. In order to understand this coupling mechanism, we utilize a modified coupled-dipole approximation (CDA)[20,21]. Similar structures were studied earlier from quantitative point of view but a rigorous analytical model has not been presented yet for such a complex multilayered structure. In this approach, the out-of-plane and in-plane coupling between (hole-disk and disk-disk) dipoles mediated by near-field and far-field radiation is calculated using the Lippmann-Schwinger equation [22-25]

$$\mathbf{E}(\mathbf{r}) = \mathbf{E}_0(\mathbf{r}) + k_0^2 \int_V \hat{G}(\mathbf{r},\mathbf{r}')\mathbf{P}(\mathbf{r}')\,d\mathbf{r}', \tag{1}$$

where $\mathbf{E}_0$ is the external electric field, $\mathbf{P}$ is the electric dipole, $k_0$ is the wavenumber in the vacuum and $\hat{G}(\mathbf{r},\mathbf{r}')$ is the Green's tensor of the nanostructure without the presence of scattering. Strong vertical hole-disk and lateral disk-disk dipolar electric field coupling dictates the present subwavelength transmission as opposed to the surface plasmon polariton (SPP) or localized surface plasmon (LSP) guided transmission through planer subwavelength hole array[26-28].

The hole and disk array in proximity (Fig. 1a) form a coupled system with two degrees of coupling, hole-disk coupling and disk-disk coupling, respectively. The disk-disk and hole-disk assisted localized surface plasmon resonance (LSPR) makes the hole-disk array more transmissive compared to a simple hole array[26-28]. The cross-sectional electric field distribution in Fig. 1b shows strong vertical hole-disk and lateral disk-disk coupling at resonance. By placing the perforated film on top of the disk-array, the weak transmitted electric field in the 3-6 µm wavelength range for the present example (Fig. 1a) excites dipolar localized surface plasmon resonance (LSPR) on the disks. Localized charges on the edges of the disks attract opposite charges on the corresponding edges of the top hole and the lateral disk array by near-field coupling as can be observed from the electromagnetic simulation in Fig. 1b. Due to the strong coupling between each disk-disk and hole-disk pair, the coupled system functions as an optical antenna array and forward scatter (transmit) enhanced electromagnetic energy as can be seen from Fig. 1a. The complete process can be understood from a 3D couple-dipole model where each element is assumed as dipole arranged in a quasi-3D periodic array. The role of coupling between different elements cannot be captured by the discrete dipole approximation (DDA) as the effect of near- and far-field coupling in the system is a major contributing factor in deciding the overall transmission[25]. In this section, we develop a couple-dipole approximation based numeric model to understand the underlying extraordinary transmission and absorption mechanism of the coupled hole-disc array and cavity-coupled hole-disk array [26-28]. According to Eq. (1), the dipolar electric field of the disk array is larger than the perforated film. This is a direct consequence of the electric dipole moment of the disk being stronger than hole at resonance due to the higher charge concentration and longer lifetime of localized surface plasmons on the isolated disks (less number of channels for radiation and resistive loss decay).

Due to the dipolar charge oscillations, the disk array re-radiates a part of this energy in all directions. Most of the radiated energy from the disk in the upward direction is reflected back from the hole array as the transmission of the hole array is very low (see Fig. 1a). This radiation is added in-phase to the radiation in the downward direction resulting in high transmission. To further understand the relation between this coupling mechanism and the extraordinary transmission of the hole-disk array, one needs to study the behavior of disk and hole arrays independently. Hole-disk combination forms a complementary system and the transmission through the hole is approximately equal to the reflection from the disk array as per the Babinet's principle [29].

The polarizabilty ($\alpha$) and the lattice contribution (S) of the circular disk array is used to estimate reflection (r) and transmission (t) coefficients of the disk array[29]. Subsequently, the reflection and transmission of the disk array is used to estimate the reflection/transmission of the hole array based on the relation, $t_{hole}(r_{hole}) = r_{disk}(t_{disk})$[29]. This is a good approximation for the sub-wavelength ($\lambda > (D, P)$) system at hand. The reflection/transmission coefficients of these independent hole and disk array are used to develop the transmission coefficient of the combined hole-disk system based on Fresnel expression of the resultant multi-layer stack. The CDA predicted transmission of the coupled hole-disk system is plotted along with results from electromagnetic simulation in Fig. 1a. The CDA prediction matches numerical simulation closely, vindicating the validity of the analytical model (the detailed CDA derivation is in the SI). The drawback of this model is different plasmon lifetimes of disk and hole which means the loss of plasmons (electron resistive loss and radiation channels) on hole-array has been underestimated which gives rise to the lower bandwidth than the experimental measurement, as can be seen from Fig. 1.a.

The relief depth and period of the hole-array plays an important role in determination of the transmission characteristics of the system. A detailed analysis of relief depth and period is presented in the supplementary information. The interaction between the two complementary elements (hole/disk) and incident electromagnetic wave is further enhanced through cavity-coupling (Fig. 2a) as can be observed in Fig. 2b. Constructive interference within the cavity excites stronger electric dipoles on the two complementary elements (see Fig. 2b-inset), resulting in 100% transmission and eventual absorption of the incident photon as the surface plasmons decay due to the resistive loss. In presence of cavity, there are two degenerate modes (electric and magnetic dipoles resonance) generate a zero slope band structure and zero group velocity ($v_g = \partial \omega / \partial k$) which enhance light-matter interaction. To further elucidate the fundamental physics behind the cavity coupling mechanism with the two complementary subwavelength metallic elements, we use the CDA method to model these interactions accurately and compare with the numerical FDTD simulations. A close correspondence is observed between the FDTD simulation and CDA prediction which supports the claim that the hole-disk, disk-disk and hole-disk-mirror interactions are predominantly responsible for the system's unique light funneling ability. The corresponding electric field distribution predicted from the FDTD simulation at resonance is shown in Fig. 2b-inset. It can be observed that the lateral dipolar coupling between discrete gold disks (disk-disk) is enhanced by more than one order of magnitude in presence of the cavity (compare Fig. 1b and Fig. 2b-inset).

The narrow bandwidth of this photon capture and subsequent absorption originates from the narrow cavity phase relation based on the quality factor (Q) of the cavity as can be observed from the FDTD simulated cavity dispersion in Fig. 3. The simulated absorption spectrum as a function of cavity length and wavelength is shown in Fig. 3 for the pattern having P = 1.140 µm,

D = 0.760 µm and RD = 280 nm. The predicted first order (m = 1) Fabry-Perot (FP) mode ($L = m\lambda/2n_{eff}$) corresponding to a simple planer cavity has been plotted on top of the FDTD simulation data in Fig. 3 (middle). A clear deviation in resonance of the cavity-coupled plasmon response from the simple FP resonance is evident due to the extra phase shift acquired by the cavity mode in presence of the disk at the localized surface plasmon resonance. This extra phase shift ($\Delta\varphi$) makes the effective thickness of the cavity longer which shifts the resonance to a longer wavelength. To quantify this effect, the phase lag from the disk array is calculated following analytical CDA approach, where the phase shift due to the disk array is used to perturb the ideal cavity phase relation resulting in a shift in the cavity resonance as can be seen in Fig. 3 (left). As expected there is a $\pi$ phase lag between the dipole array oscillator and the incident wave [30] (Fig. 3 (left)) at the LSPR wavelength (Fig. 3 (right)). Away from the plasmon resonance, the extra phase shift approaches the steady state phase due to the change in index in presence of the metallic disk array inside the cavity. The predicted resonance peak shift of the cavity-coupled hole-disk array with respect to the planer Fabry-Perot resonance for the first order mode (m = 1) calculated by CDA and numerical simulation is shown in Fig. S3. The close match between CDA and simulation vindicates the validity of the analytical phase approximation of the CDA approach. According to Fig. 3, for lower cavity thicknesses, the Fabry-Perot resonance peak falls within the LSPR bandwidth of the plasmon resonance. In addition, for these thicknesses, for the chosen relief depth (RD = 280 nm), the disk array approximately satisfies the quarter wave condition due to the additional phase shift. As a result, the system fulfills all conditions to maximize the electric field on the disk array and consequently the absorption is maximized in this region. At higher cavity thicknesses, these conditions do not satisfy simultaneously and hence the absorption goes down. However, higher order modes (m = 2 and m

= 3) of the cavity satisfy all of these conditions over the chosen cavity thickness range resulting in high absorption over the entire range as can be seen in Fig. 3.

Due to the geometrical origin, the response can be flexibly tuned by changing dimensions (period, diameter, distance between two elements and cavity phase) of the coupled system (see Supplemental Information). The proposed system is fabricated following a simple large area nanoimprinting technique[17]. A polydimethylsiloxane (PDMS) stamp is embossed against a photoresist (SU-8) layer spun coated on a glass substrate coated with optically thick (200 nm) gold mirror. The blanket deposition of thin layer (30 nm) of gold completes the simple fabrication process. The deposited gold forms the top perforated hole pattern on the raised region of the polymer imprint and the bottom disk array in the recessed region. Figure 4a shows the scanning electron microscope (SEM) images of five such representative systems with varied hole/disk diameter for constant D/P = (0.60-0.66). The corresponding optical photon capture via absorption measurements using a microscope-coupled FTIR (Hyperion 1000-Vertex 80, Bruker Inc.) along with simulation predictions appear in Fig. 4b. As predicted, near 100% of the incident radiation is captured (measured as absolute absorption). We have noticed about 3% variation in absorption and 5% variation in spectral peak location across a batch of samples made from five different imprinting stamps as can be seen in Supplementary Fig. S4. Such variation originated primarily due to the manual imprinting process and can be reduced further with the implementation of an automated imprinting system.

The proposed work demonstrates a unique cavity phase driven extraordinary transmission through a subwavelength complementary aperture pair where previously observed extraordinary transmission is not possible due to the absence of natural plasmon resonance at that wavelength range. The cavity phase driven tunable photon capture opens up a new way of enhancing light-matter interactions for practical applications like frequency selective infrared detection, bio-sensing and light harvesting.

**Acknowledgments:**

This work at University of Central Florida was supported by the Florida Space Institute/NASA grant no. 63019022 and Northrop Grumman Corporation grant no. 63018088.

**Author contributions:**

D.C. conceived the idea. D.C designed and S.M. performed the experiments. A.S. provided technical guidance and developed analytical models.  S.M. and A.S. analyzed and simulated the data. A.S., S.M. and D.C. co-wrote the paper.

# Figures

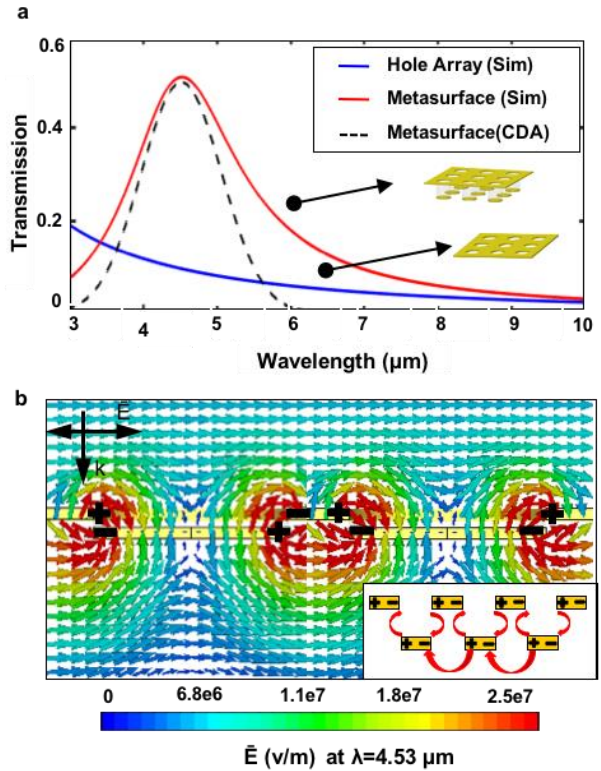

**Figure 1:** (a) Predicted light transmission through the subwavelength hole (blue) and hole-disk (red) array of period P = 1.14 µm, diameter D = 0.76 µm and relief depth RD = 74 nm along with the analytical coupled dipole approximation (CDA) prediction (black-dotted) have been overlaid for comparison. (b) FDTD predicted cross-sectional electric field distribution showing vertical hole-disk and lateral disk-disk coupling at resonance ($\lambda$ = 4.53 μm).

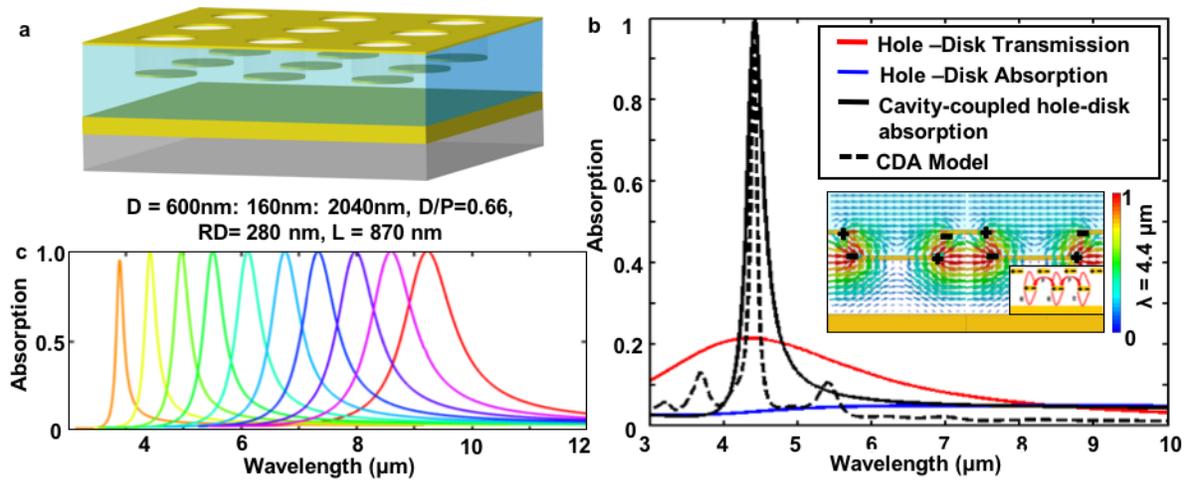

**Figure 2:** (a) The schematic of the cavity-coupled hole-disk system. (b) Predicted light absorption (blue) and transmission (red) through the subwavelength hole-disk array of period P = 1.14 µm, diameter D = 0.76 µm and relief depth RD = 280 nm. The measured (black) photon capture efficiency of the coupled system along with the analytical coupled dipole approximation (CDA) prediction (black-dotted) have been overlaid for comparison for cavity thickness L = 0.87 µm. Inset: Computed electric field distribution inside the cavity-coupled absorber. (c) Predicted light absorption of the cavity-coupled hole-disk array with different diameters (D = 600 nm: 160 nm: 2040 nm, D/P=0.66), relief depth RD = 280 nm and cavity thickness L= 870 nm.

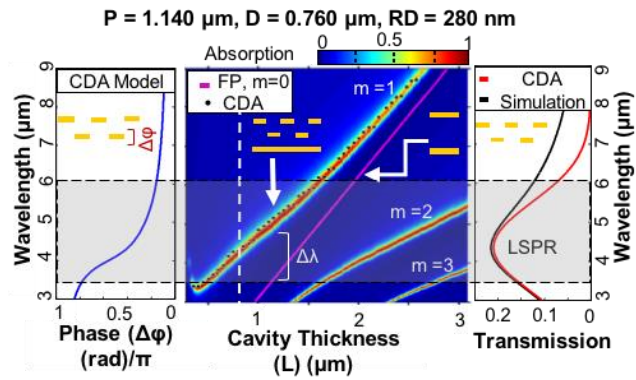

**Figure 3:** FDTD predicted change in absorption as a function of wavelength and cavity length of the system with period P = 1.14 µm, diameter D = 0.76 µm and relief depth RD = 280 nm. The extra phase due to the presence of the coupled disk array inside the cavity is calculated from CDA method (left) and the corresponding LSPR (Transmission of the bare hole-disk) calculated from simulation and CDA method is shown in (right).

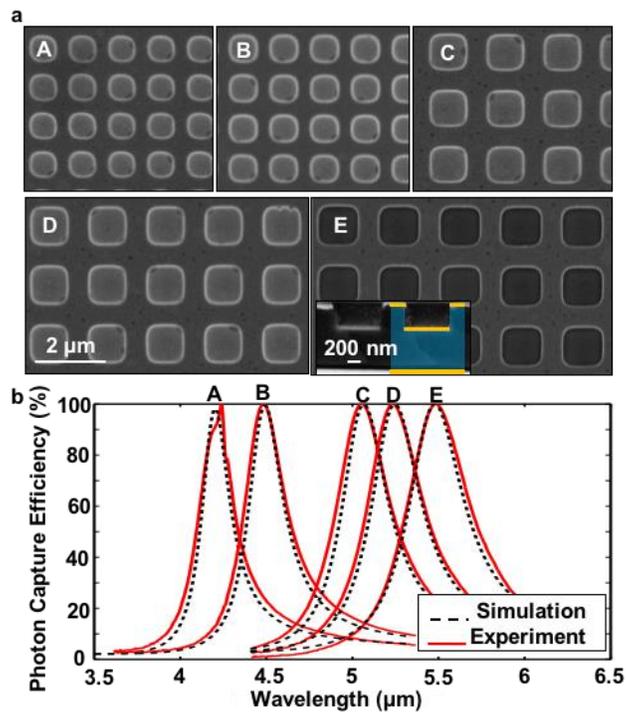

**Figure 4:** (a) Scanning electron microscope (SEM) images for five cavity-coupled absorbers with periods (P = 1.1, 1.14, 1.6, 1.66 and 1.74μm in frames A-E) with diameter to period ratio is kept constant (D/P = 0.6). Inset shows the cross section of the quasi-3D structure. (b) Measured photon capture/absorption spectra for the periods shown in part (a) along with simulated spectra for each period.

# Supplementary Information

**Couple-dipole approximation (CDA):**

In CDA approach, firstly polarizability of the disk array is calculated as [1]:

$$\alpha_{cluster} = \frac{\alpha_S}{1-\alpha_S S}, \tag{S1}$$

where $\alpha_S$ is the is the polarizability for a single disk and $S$ is lattice contribution

$$S = \sum_{j \neq i}[\frac{(1-ikr_{ij})(3\cos^2\theta_{ij}-1)e^{ikr_{ij}}}{r_{ij}^3} + \frac{k^2\sin^2\theta_{ij}e^{ikr_{ij}}}{r_{ij}}]. \tag{S2}$$

The polarizability of a generalized ellipsoidal nanoparticle is

$$\alpha_S = \varepsilon_0 V \frac{\varepsilon - \varepsilon_m}{\varepsilon + L_a(\varepsilon - \varepsilon_m)}, \tag{S3}$$

where $\varepsilon$ and $\varepsilon_m$ are the dielectric function of the metal and surrounding medium, respectively. $V$ is the volume and $L_a$ is the shape factor of the ellipsoid is defined as:

$$L_a = \frac{abc}{2}\int_0^\infty \frac{dq}{(a^2+q)\sqrt{(a^2+q)(b^2+q)(c^2+q)}} \tag{S4}$$

where $a$ is the diameter of the ellipsoid along the polarization direction, $b$ and $c$ are the diameters along other two dimensions.

The reflection coefficient of the disk array is [2]

$$r_{disk} = \frac{\pm iG}{\alpha^{-1}-S}, \tag{S5}$$

where

$$G = \frac{2\pi k}{A}\begin{cases}\cos^{-1}\theta, s-polarization \\ \cos\theta, p-polarization\end{cases} \tag{S6}$$

$k$ is wavenumber and $A$ is the area of unit cell. Positive and negative signs stand for s and p-polarization.

The transmission coefficient of this disk array is $t_{disk} = 1 \pm r_{disk}$.

The reflection/transmission of the disk array is used to estimate the reflection/transmission of the hole-array based on the approximate relation,

$$\begin{cases} r_{hole} = t_{disk} \\ t_{hole} = r_{disk} \end{cases}. \tag{S7}$$

The reflection and transmission coefficients of a thin layer on top of a substrate, are estimated as

$$r_{tot} = r + \frac{(1 \pm r)^2 r^0 e^{2i\beta}}{1 - r^0 r e^{2i\beta}}, \tag{S8}$$

and

$$t_{tot} = \frac{(1 \pm r) t^0}{1 - r^0 r e^{2i\beta}}. \tag{S9}$$

In these equations, positive and negative signs stand for s and p-polarization. $r$ is the reflection coefficient of the thin film. $r^0$ and $t^0$ are reflection and transmission coefficients of the substrate. In addition, $\beta$ is the propagation phase in the thin-film

$$\beta = \beta_r + i\beta_i = \sqrt{\varepsilon_{spacer}} k_0 d. \tag{S10}$$

To obtain hole-disk system reflection and transmission analytically, Eq. (S1)-(S7) were used with multiple reflection formula Eq. (S8)-(S10), as:

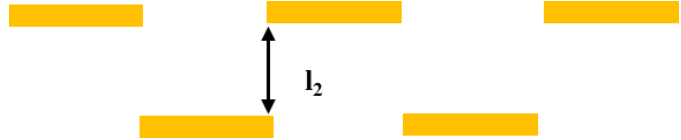

$$r_{disk+hole} = r_h + \frac{(1 + r_h)^2 r_d e^{i(2nk_0 l_2)}}{1 - r_d r_h e^{i(2nk_0 l_2)}}, \tag{S11}$$

$$t_{disk+hole} = \frac{(1 + r_h) t_d}{1 - r_d r_h e^{i(2nk_0 l_2)}}. \tag{S12}$$

In the CDA computation, a large number of particles (10,000) were considered in order to account for the far-field interaction between the holes and the disks.

In the cavity coupled case the hole-disk formulation is modified to account for the reflection from the back mirror and the round trip phase in Eq. (S12) as:

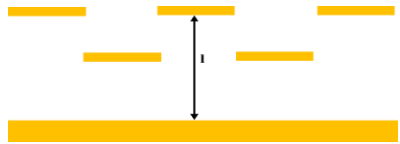

$$r_{tot} = r_{Au} + \frac{(1+r_{Au})^2 r_{disk+hole} e^{i(2n_{eff} k_0 l)}}{1 - r_{disk+hole} r_{Au} e^{i(2n_{eff} k_0 l)}}, \qquad (S13)$$

where $l$ is the distance between gold back mirror and the hole at top.

The effective index in Eq. (S13) was calculated from Maxwell-Garnet theorem. The penetration of EM wave inside the gold is wavelength dependent. So, by using effective medium theory, we calculated effective refraction index of the spacer [3-6]. We considered different materials in z and r directions. To derive the effective refractive index, we divided the system into three regions as schematically shown below:

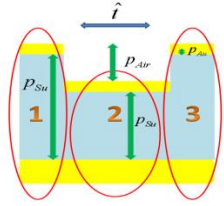

$$\langle D \rangle = \varepsilon_{eff} \langle E \rangle, \qquad (S14)$$

$$\varepsilon_{1eff} = \varepsilon_{3eff} = \frac{2 p_{Au} \varepsilon_{Au} + p_{Su} \varepsilon_{Su}}{2 p_{Au} + p_{Su}}, \qquad (S15)$$

and

$$\varepsilon_{2eff} = \frac{p_{Air} \varepsilon_{Air} + 2 p_{Au} \varepsilon_{Au} + p_{Su} \varepsilon_{Su}}{p_{Air} + 2 p_{Au} + p_{Su}}. \qquad (S16)$$

In these regions, because the polarization is parallel to the surface, so the electric fields in different regions are equal ($E_{1t} = E_{2t}$). In addition, $p_{Au}$ is equal to skin depth of the gold. Now three regions are defined by effective indices as illustrated below:

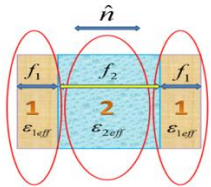

The effective permittivity can be estimated by considering $D_{1n} = D_{2n}$

$$\varepsilon_{eff} = \frac{2f_1\varepsilon_{1eff}E_1 + f_2\varepsilon_{2eff}E_2}{2f_1 E_1 + f_2 E_2}, \tag{S17}$$

Now, solving Eq. (S16) and (S17), one can derive the effective permittivity as

$$\varepsilon_{eff} = \frac{2f_1 + f_2}{\dfrac{2f_1}{\varepsilon_{1eff}} + \dfrac{f_2}{\varepsilon_{2eff}}}. \tag{S18}$$

With this coupled system, a good agreement between CDA and simulation can be observed (fig. 1 of the main manuscript). As an example for a chosen hole-disk period, P = 1.140 µm, diameter, D = 0.760 µm and relief depth RD = 74 nm ($n_{air}$ = 1, $n_{SU-8}$ = 1.56) the effective permittivity of $\varepsilon_{eff}$ = 1.42 is predicted from Eq. (S18).

**Analysis of effect of disk array position- in the cavity and from the hole array**

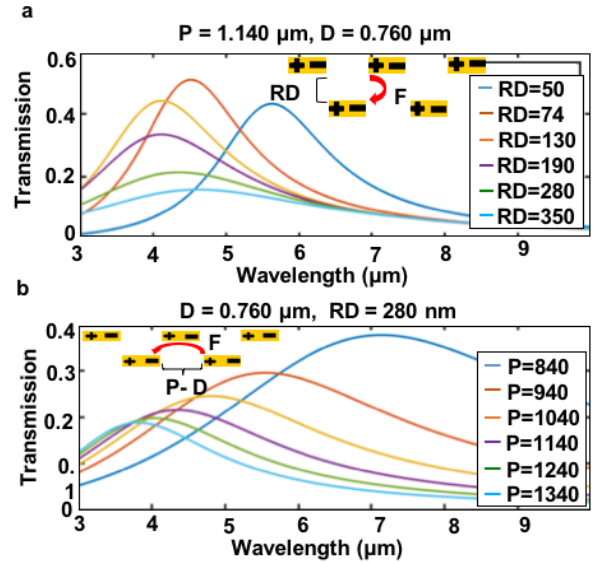

**Fig. S1.** Plot of transmission through hole-disk array for (a) varying relief depth (50-350nm) and constant period(1.14µm) and diameter (0.76µm). (b) for a fixed diameter (760nm) and relief depth (280nm) and varying period (840nm-1430nm).

The cross-sectional electric field distribution in Fig. 1b shows strong vertical hole-disk and lateral disk-disk coupling at resonance which strongly depends on the geometrical spacing between the two complementary elements as can be seen in Fig. S1. The overall response of the system can be tuned by manipulating the coupling between hole-disk and disk-disk by changing dimension or array spacing. By changing the relief depth (RD), two different regimes (phases) in hole-disk coupling are observed. They manifest as blue and then red shift of the transmission peak as a function of increase in RD as shown in electromagnetic simulation in Fig. S1a (period and diameter, D are kept constant at 1.14 µm and 0.76 µm to ensures constant disk-disk coupling). This indicates the presence of a phase transition and critical point $(RD_c)$ [7-9] in hole-disk coupling. By decreasing RD, the electrostatic and radiative coupling between hole and disk is increases (Eq. 1) which in turn makes them to exert a larger electric field on each. This enhanced field is opposite to the restoring field of the surface charges on the disk resulting in charge resonance. Due to the decrease in the effective electric field, the restoring force spring constant $k_{eff}$ also decreases, which redshifts the transmission peak as per the relation $\omega_{res} \propto \sqrt{k_{eff}}$.

Moreover, the decrease in RD from the critical value, increases the opposing electric field on the holes resulting in reduction in the effective electric dipole on the disk. Hence, according to the Eq. 1, the total transmission of the disk and hole arrays decreases. Additionally, by increasing RD from $RD_c$, the coupling between hole-disk becomes weaker as a result the transmission through the hole and total transmission of the structure diminishes. Due to the reduction in transmission, the localized charges on the disk decrease. This effect diminishes the restoring force, leading to lower spring constant $k_{eff}$ and a red shift for the transmission peak. For a constant hole-disk coupling (constant RD), the extraordinary transmission peak location can be tuned over a wide range by manipulating disk-disk coupling by means of changing the diameter and period of the system. For this study a relatively large RD = 280 nm is used for the ease of actual fabrication of this structure at a later stage because a small RD increases the possibility of destroying the isolation between the hole and disk system during metal deposition. On the other hand, as the period increases, the distance between elements also increases, and hence, the opposite electric fields on the elements that originate from neighboring localized charges decrease. It means that the effective electric field on the surface and consequently the effective spring constant, $k_{eff}$ is increased which results in blue shift in the spectrum, as shown in Fig. S1b. In Eq. (1), $|\mathbf{P}| = qd$ is electric dipole moment, $q$ is the charge (which depends on the electric field strength) and $d$ is the distance between the charges (which is approximately equal to the diameter for localized surface plasmon (LSP)). As the localized surface plasmon resonance (LSPR) wavelength on the disk is inversely proportional to the diameter, an expected red shift is observed with increase in hole/disk diameter for a constant period as seen in Fig. S2a. Also, for the same case, as the hole diameter is increased, the electric dipole ($\mathbf{P}$) gets stronger and as a result, the transmission is increased as well. The LSPR in the coupled system experiences red and blue shifts by increasing diameter and period, respectively. With the increase in the period, electric field coupling between neighboring disks/holes decreases. Hence, the effective spring constant $k_{eff}$ and the resonance wavelength decrease with the increase in period in an implicit manner. A closed form expression that relates period with effective spring constant does not exists for 2D array due to complexities in multi-modal and multi-dimensional interactions.

## LSPR vs. diameter

The LSPR resonance frequency depends on the force between localized charges

$$\omega_{res} \propto \sqrt{k}, \tag{S19}$$

where $k$ is the spring constant given by

$$k = -\frac{\partial F}{\partial x}\bigg|_{x=D}, \tag{S20}$$

where $F$ is the total coulomb force between positive and negative charges which depends on the number of separating charges (LSP's), $N$, and distance between them which approximately equals to diameter, $D$,

$$F \propto \frac{N}{D^2}. \tag{S21}$$

In 2D, the number of particles on the surface depends on $D$. Combining Eq. (S19)-(S21) gives

$$N_{2D} \propto D \Rightarrow \omega_{res} \propto \frac{1}{D}. \tag{S22}$$

From Eq. (S22) it can be concluded that the LSPR red shifts with the increase in diameter of the particle. This relation between diameter and resonance frequency is not exact, but it gives us estimation for the behavior of this kind structures, as shown in Fig. S2 deriving by means of 3D full-vectorial electromagnetic simulation and CDA approach.

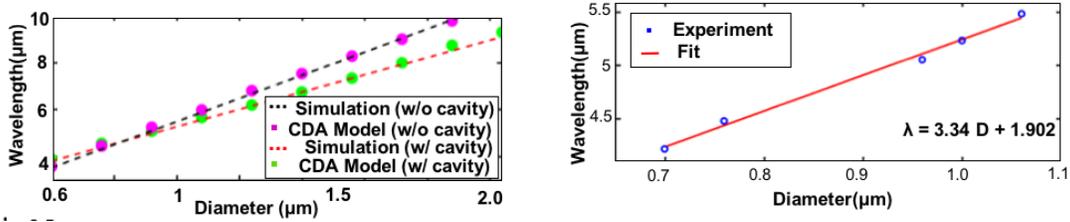

**Fig. S2.** Left: Comparison of FDTD simulation and CDA model in extraction of the peak absorption wavelength as a function of diameter with and without cavity. The relief depth (RD=0.28µm) and cavity height (L=0.87µm) are kept constant and diameter is varied from 600nm to 2040 nm in steps of 160nm keeping D/P ratio as 0.65. Right: The experimental realization of peak absorption wavelength as a function of diameter with D/P~(0.6-0.65).

## Cavity phase evaluation

By using Eq. (S5), we calculated the added phase ($\theta_{LSP}$) of the transmitted electromagnetic wave according to the incident wave ($t = |t|e^{i\theta_{LSP}}$). After that by using Drude model for gold, we calculated the phase of reflected waves, by using Fresnel equations ($r = \dfrac{\sqrt{\varepsilon_{Au}} - \sqrt{\varepsilon_{Su}}}{\sqrt{\varepsilon_{Au}} + \sqrt{\varepsilon_{Su}}}$); Actually this phase is $\varphi_{ref}$ ($r = |r|e^{i\varphi_{ref}}$) which is $\pi$ for perfect conductors. Now, by adding $\theta_{LSP}$, $\varphi_{ref}$ and the optical path length inside the cavity ($\gamma = 2(\dfrac{2\pi n_{su}}{\lambda})l_{su}$), we can calculate the total phase which EM wave acquires inside the cavity. After that, by comparing this phase with the phase shift of a regular Fabry-Perot resonator, we obtained the effective thickness ($l_{eff}$) of the structure

$$(\varphi_{tot} = \theta_{LSP} + \varphi_{ref} + \gamma = \varphi_{ref} + 2(\dfrac{2\pi n_{eff}}{\lambda})l_{eff}).$$

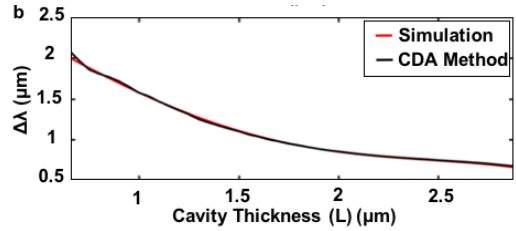

**Fig. S3**. Change in the resonance wavelength due to the presence of disk array in the cavity as a function of cavity thickness. The position of disk array is kept constant at a distance of 280nm from the hole-array. FDTD and CDA results are plotted for comparison.

**Absorber sample statistics**

To understand the variation in optical antenna absorption, samples made from different imprinting stamps were studied. We have noticed about 3% variation in absorption and 5% variation in spectral peak location across a batch of samples made from 5 different imprinting stamps as can be seen in Fig. S3. Such variation originated primarily due to the manual imprinting process. We believe an automated imprinting technique with finer control will reduce these minor variations to even small numbers.

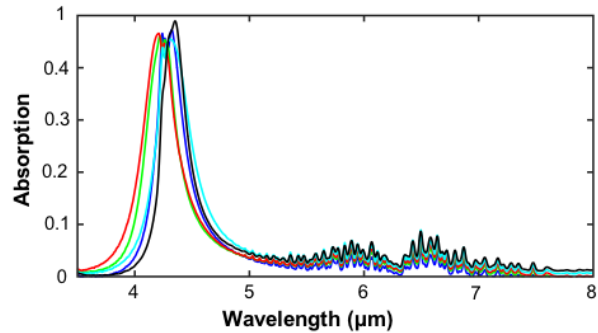

**Fig. S4.** Absorber sample statistics. We have noticed about 3% variation in absorption and 5% variation in spectral peak location across a batch of samples made from 5 different imprinting stamps. The parameters are- period = 1140nm, diameter = 760nm, relief depth = 280nm and cavity thickness = 870nm.